\documentclass{amsart}
\usepackage{graphicx}
\usepackage{epsf}

\usepackage{amsmath, amssymb,latexsym,float,algorithm,algorithmic}
\usepackage{enumerate}

\theoremstyle{definition}

\theoremstyle{remark}

\numberwithin{equation}{section}

\newcommand{\R}{\text{\fontshape{n}\selectfont I\kern-.42exR}}

\newcommand{\1}{\text{\fontshape{n}\selectfont 1\kern-.56exl}}

\begin{document}

\centerline{\Large\it University of Edinburgh}

\title[UV-suppressed quasi-optimal domain wall fermions]
{A note on ultraviolet suppressed quasi-optimal domain wall fermions}

\author{Artan Bori\c{c}i}



\date{October 2002}
\maketitle
\begin{abstract}
In a recent work Chiu proposed to modify domain wall fermions
that allow in an optimal way fewer number of flavors than in
the standard case. This is done using a variant of doamin wall
fermions, the so-called truncated overlap fermions.
In this note I discuss the possibility
to implement his proposal for the original variant of domain wall
fermions.
I make also some remarks on dynamical simulations with
ultraviolet suppressed domain wall fermions.
\end{abstract}
\pagebreak

\subsection*{Domain wall and overlap fermions}

I will briefly reveiw the definitions of domain wall fermions
\cite{Kaplan,Furman_Shamir}.
Let $N$ be the size of the extra dimension, $D_W$ the Wilson-Dirac
operator, and $m$ the bare fermion mass. Then, the domain wall
fermion action is given by:
\begin{equation}
S_{DWF} := \bar{\Psi} {\mathcal D} \Psi = \sum_{i=1}^{N}
  \bar{\psi}_i [a_5(D^{||}-\1)\psi_i + P_{+}\psi_{i+1} + P_{-}\psi_{i-1}]
\end{equation}
together with boundary conditions:
\begin{equation}
\begin{array}{l}
P_{+}(\psi_{N+1} + m \psi_1) = 0 \\
P_{-}(\psi_0 + m \psi_N) = 0
\end{array}
\end{equation}
Here $\mathcal{M}$ is the five-dimensional fermion matrix of the
theory and $D^{||} = M - D_W$ with $M \in (0,2)$ being a mass parameter.
The alternative theory of chiral fermions on the lattice is
formulted by \cite{Narayanan_Neuberger}. In the domain wall setting
this theory can by eqiuvalently fomulated using truncated
overlap fermions \cite{Borici_TOV,Borici_MG}:
\begin{equation}
S_{TOV} := \bar{\Psi} {\mathcal D}_{TOV} \Psi = \sum_{i=1}^{N}
  \bar{\psi}_i [a_5(D^{||}-\1)\psi_i
  + (a_5D^{||}+\1)P_{+}\psi_{i+1} + (a_5D^{||}+\1)P_{-}\psi_{i-1}]
\end{equation}
with the same boundary conditions as above.
In both cases the four dimensional lattice spacing is set to one.

It is well-known that domain wall fermions in both variants can be
dimensionally reduced to give an effective theory of four dimensional
chiral fermions \cite{Neuberger_DWF,Kikukawa_Noguchi,Borici_99}.
The Dirac operator of the theory is given by the Neuberger overlap
operator \cite{Neuberger1}:
\begin{equation}\label{overlap_op}
D = \frac{1+m}{2}~\1 - \frac{1-m}{2} \gamma_5 ~\text{sgn}(H)
\end{equation}
where $H = \gamma_5 D^{||}$ in case of
truncated overlap fermions with transfer matrix \cite{Borici_TOV}:
\begin{equation}
T = \frac{\1 + a_5H}{\1 - a_5H}
\end{equation}
For domain wall fermions it was shown that \cite{Borici_99}:
\begin{equation}
\mathcal{H} = \gamma_5 \frac{D^{||}}{2 - a_5D^{||}} = H \frac{1}{2 - a_5D^{||}}
\end{equation}

Recently, Chiu proposed to use a transfer matrix which is depends on
the 5D time slice \cite{Chiu2002}:
\begin{equation}\label{optimal_T}
T_i = \frac{\1 + \omega_i a_5H}{\1 - \omega_i a_5H}
\end{equation}
where $\omega_i, i=1,\ldots,N$ are positive coefficients and
$a_5 = 1$. These
coefficients are chosen such as to correspond to the optimal
Chebyshev rational approximation of the sign function, the
Zolotarev approximation \cite{Zolotarev}. In his paper, Chiu
showed how to implement this idea in case of truncated overlap
fermions. Below I discuss Chiu's idea in the case of
domain wall fermions.

\subsection*{Quasi-optimal domain wall fermions}
I postulate the optimal domain wall fermion action given by:
\begin{equation}
  \bar{\Psi} {\mathcal D} \Psi = \sum_{i=1}^{N}
  \bar{\psi}_i
  [(\omega_ia_5D^{||}-\1)\psi_i + P_{+}\psi_{i+1} + P_{-}
  \psi_{i-1}]
\end{equation}
To show that such domain wall fermions are optimal it is sufficient
to calculate tranfer matrices $T_i, i=1,\ldots,N$ from $\mathcal M$
and find that they are in the optimal form (\ref{optimal_T}). The fermion
matrix can be written as an $N \times N$ block partition along the
fifth dimension:
\begin{equation*}
\mathcal M =
\begin{pmatrix}
\omega_1a_5D^{||}-\1 & P_+                  &        & -mP_-                \\
P_-                  & \omega_2a_5D^{||}-\1 & \ddots &                      \\
                     & \ddots               & \ddots & P_+                  \\
-mP_+                &                      & P_-    & \omega_Na_5D^{||}-\1 \\
\end{pmatrix}
\end{equation*}
Multiplying from the right by the permutation matrix:
\begin{equation*}
\begin{pmatrix} P_+ & P_- &        &     \\
                    & P_+ & \ddots &     \\
                    &     & \ddots & P_- \\
                P_- &     &        & P_+ \\
\end{pmatrix}
\end{equation*}
I obtain:
\begin{equation*}
\gamma_5
\begin{pmatrix}
(\omega_1 a_5HP_+ - \1)(P_+ - mP_-) & \omega_1 a_5HP_- + \1 &  & \\
                                    & \omega_2 a_5HP_+ - \1 & \ddots &\\
                                    & & \ddots & \omega_{N-1} a_5HP_- + \1 \\
(\omega_N a_5HP_- + \1)(P_- - mP_+) & &        & \omega_N     a_5HP_+ - \1 \\
\end{pmatrix}
\end{equation*}
Further, multiplying this result from the left by the inverse
of the diagonal matrix:
\begin{equation*}
\begin{pmatrix}
\omega_1 a_5HP_+ - \1 &                       &  & \\
                      & \omega_2 a_5HP_+ - \1 &  & \\
                      &                       &  \ddots & \\
                      &                       &  & \omega_N a_5HP_+ - \1 \\
\end{pmatrix}
\end{equation*}
I get:
\begin{equation}
\mathcal{T}(m) :=
\begin{pmatrix} P_+ - mP_-       & -T_1 &        &          \\
                                 & \1   & \ddots &          \\
                                 &      & \ddots & -T_{N-1} \\
                -T_N(P_- - mP_+) &      &        & \1       \\
\end{pmatrix}
\end{equation}
with transfer matrices being defined by:
\begin{equation*}
T_i = \frac{\1}{\1 - \omega_i a_5HP_+} (\1 + \omega_i a_5HP_-),
\text{\hspace{1cm}}i=1,\ldots,N
\end{equation*}
By requiring transfer matrices being in the form:
\begin{equation*}
T_i = \frac{\1 + \omega_i a_5\mathcal{H}}
           {\1 - \omega_i a_5\mathcal{H}}
\end{equation*}
it is easy to see that \cite{Borici_99}:
\begin{equation}
\mathcal{H} = H \frac{1}{2 - \omega_i a_5D^{||}}
\end{equation}

Note that $\mathcal{H}$ depends in the time slice $i$.
By ignoring the denominator the resulting transfer matrix
will be of the optimal form (\ref{optimal_T}). This may lead
to suboptimal approximation of the sign function and must be
studied numerically.

Finally to derive the four dimensional Dirac operator one has to
compute:
\begin{equation*}
\det D^{(N)} = \frac{\det\mathcal{T}(m)}{\det\mathcal{T}(1)}
\end{equation*}
It is easy to show that the determinant of the $N \times N$
block matrix $\mathcal{T}(m)$ is given by:
\begin{equation*}
\det\mathcal{T}(m) = \det[(P_+ - mP_-) - T_1 \cdots T_N (P_- - mP_+)]
\end{equation*}
or
\begin{equation*}
\det\mathcal{T}(m) =
 \det[  \frac{1 + m}{2} \gamma_5 (\1 + T_1 \cdots T_N)
      + \frac{1 - m}{2}          (\1 - T_1 \cdots T_N)]
\end{equation*}
which gives:
\begin{equation*}
D^{(N)} = \frac{1 + m}{2} \1 + \frac{1 - m}{2} \gamma_5
        \frac{\1 - T_1 \cdots T_N}{\1 + T_1 \cdots T_N}
\end{equation*}
In the large $N$ limit one gets the Neuberger operator (\ref{overlap_op}).

\subsection*{Ultraviolet suppressed domain wall fermions}

One difficulty of dynamical simulations with domain wall fermions
is the presence of many heavy fermion modes which are simulated
together with the interesting light fermion. The problem, generelly
knwon as the ultraviolet slowing down becomes more severe in this case.
A possible solution is to subtract the contribution of heavy modes in the
first place \cite{RBC_callab}.

Here I discuss another solution which is based on the recent proposal
of ultraviolet suppressed fermions \cite{Borici_UVSFa}. This proposal
simply states that the lattice Dirac operator can be modified to give:
\begin{equation}\label{D_operator}
D = \frac{\mu}{a_5} \Gamma_5 \tanh \frac{a_5 \Gamma_5 D_{\text{DWF}}}{\mu}
\end{equation}
where $D_{\text{DWF}}$ is the input domain wall operator, $a$ the
lattice spacing and $\mu > 0$ is a dimensionless parameter
and $\Gamma_5 = \gamma_5 R_5$. $R_5$ is the reflection
operator with respect to the 5-dimensional midpoint \cite{Blum_et_al}.
The Dirac operator converges to the input operator in the contimuum limit
and is local and unitary as shown in detail in
\cite{Borici_UVSFa}. The input theory is also recovered in the limit
$\mu \rightarrow \infty$. For $\mu \rightarrow 0$ one has
$D \rightarrow \mu$, i.e. a quenched theory.

Note that the parameter $\mu$ has a special role in this theory.
Fixing $\mu$ at small values, typically $\mu \sim 1$ one can achieve
the desired suppression of heavy fermion modes of the input domain
wall operator. This way, one localizes the interesting light mode.

At the first sight, it looks that the introduction of a matrix function
can lead to potentially difficult computationas. In fact, it can be
shown that using Lanczos based methods and multi-shift conjugate gradient
iterations the computational overhead is small \cite{Borici_UVSFb}.

The benefit using the ultraviolet suppressed operator is hence twofold:
one stays within the conventional domain wall formalism of chiral fermions
and accelerates dynamical simulations by suppressing both heavy and ultraviolet
modes.

\subsection*{End note}

This work has been prepared during October 2002 when additions in the
paper of {Chiu2002} had not already appeared. It is nice to observe that
his numerical computations support the proposal of using quasi-optimal
domain wall fermions.

\end{document}